\documentclass[ reprint,amsmath,amssymb,aps,twocolumn,superscriptaddress,longbibliography]{revtex4-1}

\usepackage[nopar]{lipsum}
\usepackage{graphicx}
\usepackage{dcolumn}
\usepackage{bm}
\usepackage[capitalize]{cleveref}
\let\revappendix\appendix

\usepackage{siunitx}

\begin{document}

\preprint{APS/123-QED}

\title{
Gate-defined superconducting channel in magic-angle twisted bilayer graphene
}

\author{Giulia Zheng}
\author{El\'ias Portol\'es}
\author{Alexandra Mestre-Tor\`a}
\author{Marta Perego}
\affiliation{Laboratory for Solid State Physics, ETH Zurich,~CH-8093~Zurich, Switzerland}
\author{Takashi Taniguchi}
\author{Kenji Watanabe}
\affiliation{Research Center for Materials Nanoarchitectonics, National Institute for Materials Science,  1-1 Namiki, Tsukuba 305-0044, Japan}
\author{Peter Rickhaus}
\author{Folkert K. de Vries}
\affiliation{Laboratory for Solid State Physics, ETH Zurich,~CH-8093~Zurich, Switzerland}
\author{Thomas Ihn}
\author{Klaus Ensslin}
\affiliation{Laboratory for Solid State Physics, ETH Zurich,~CH-8093~Zurich, Switzerland}
\affiliation{Quantum Center, ETH Zurich,~CH-8093 Zurich, Switzerland}
\author{Shuichi Iwakiri}
\email{siwakiri@phys.ethz.ch}
\affiliation{Laboratory for Solid State Physics, ETH Zurich,~CH-8093~Zurich, Switzerland}

\date{\today}

\begin{abstract}
Magic-angle twisted bilayer graphene (MATBG) combines in one single material different phases like insulating, metallic and superconducting. These phases and their in-situ tunability make MATBG an important platform for the fabrication of superconducting devices. We realize a split gate-defined geometry which enables us to tune the width of a superconducting channel formed in MATBG. We observe a smooth transition from superconductivity to highly resistive transport by progressively reducing the channel width using the split gates or by reducing the density in the channel.
Using the gate-defined constriction, we control the flow of the supercurrent, either guiding it through the constriction or throughout the whole device or even blocking its passage completely. This serves as a foundation for developing quantum constriction devices like superconducting quantum point contacts, quantum dots, and Cooper-pair boxes in MATBG.
\end{abstract}
\maketitle

\section{Introduction}
Magic-angle twisted bilayer graphene (MATBG) offers highly-tunable quantum states \cite{cao_unconventional_2018,yankowitz_tuning_2019}, which have enabled a novel class of gate-defined superconducting nanodevices such as Josephson junctions \cite{de_vries_gate-defined_2021,rodan-legrain_highly_2021,diez-merida_symmetry-broken_2023}, SQUIDs \cite{portoles_tunable_2022}, and ring geometries showing Little--Parks oscillations \cite{Iwakiri2023-gl}.
All these devices rely on defining interfaces between resistive and superconducting phases. In a gate-defined narrow MATBG constriction, the interface between the superconducting and resistive phases can be continuously controlled and studied.
In Bernal bilayer graphene, a constriction (e.g. a quantum point contact \cite{Overweg2018,Nakaharai2011}) can be formed due to its gate-tunable band gap of a few 100 meV \cite{Ohta2006,Oostinga2008}. However, in the case of MATBG, the size of the band gap between the flat band and the dispersive band is one order of magnitude smaller \cite{rodan-legrain_highly_2021}. Therefore it is challenging to form a similar constriction and the
question arises, whether it is possible at all to realize a narrow superconducting channel by tuning the width of confining resistive states.

In this work, we demonstrate a gate-defined superconducting channel in MATBG. We observe a smooth transition from supercurrent to highly resistive transport (pinch-off) by progressively reducing the channel width using split gates or by reducing the density in the channel using the channel gate. We find that the supercurrent in the constriction can be turned on and off. These results will serve as the foundation for developing constriction-based devices like superconducting quantum point contacts, quantum dots, or Cooper-pair boxes in MATBG.

\section{Device and bulk characterization}
\begin{figure}[h]
\centering
\includegraphics{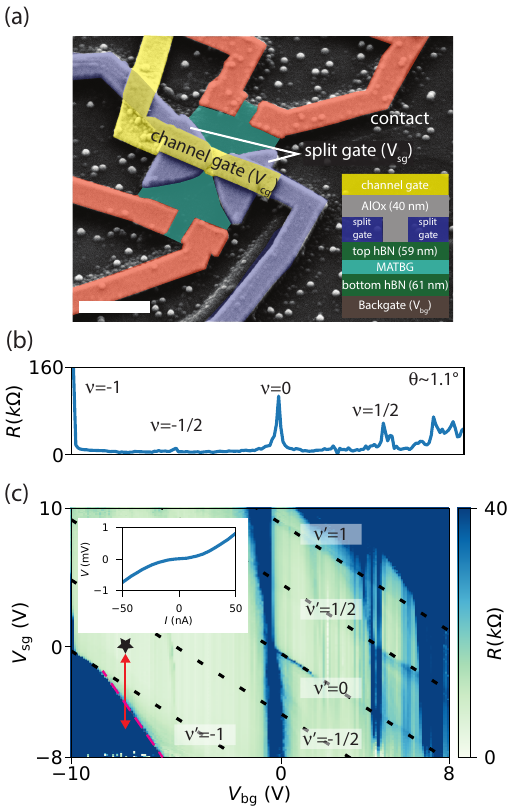}
\caption{(a) False colored SEM picture of the sample. The white scale bar at the bottom left indicates \SI{1}{\micro m}. MATBG (light green), contacts (orange), split gates (blue), and channel gate (yellow) are shown. Inset: Schematic of the stack (thickness do not scale). (b) Characterization of the device with $V_\textrm{cg}=0$. Line-cut at $V_\textrm{sg}=0$ with annotations of filling factor $\nu=-1, -1/2, 0, 1/2,$ $\textrm{and}$ $1$ states. (c) Mapping of the resistance as a function of $V_\textrm{sg}$ and $V_\textrm{bg}$. Black dashed lines labeled $\nu'=-1, -1/2, 0, 1/2,$ indicate filling factors in the split-gated area. The pink dashed line highlights the transition from low to high resistance in the hole side. The inset shows the superconducting current--voltage characteristic taken at $V_\textrm{sg}=0$ and $V_\textrm{bg}=\SI{-7.44}{V}$ (indicated by a black star). The red double-sided arrow is the gate voltage range in which Fig.~2(a) is taken.}
\label{Sample}
\end{figure}
The fabrication starts by encapsulating MATBG (light green) between two hBN layers of thickness \SI{59}{nm} and \SI{61}{nm}(dark green). Additionally a graphite layer (dark gray) is picked up, serving as the back gate. The lateral device layout is shown in Fig.~\ref{Sample}(a).
To form electric contacts (orange in the figure), we etch the top hBN using reactive ion etching (RIE) and evaporate Cr(\SI{5}{nm})/Au(\SI{65}{nm}).
On top of the top hBN, we evaporate Cr(\SI{5}{nm})/Au(\SI{15}{nm}) split gates (blue) with a gap of \SI{150}{nm}. After depositing a \SI{40}{nm} layer of aluminum oxide, an additional channel gate (yellow) with a width of \SI{400}{nm} (in the direction of current flow) is deposited, covering the region in-between the split gates.
Our gate configuration defines three different areas in the MATBG mesa: the leads shown in light green and tuned only by the back gate, the split gated region in blue, tuned by split gates and back gate together, and the channel in between the split gates and below the channel gate. Since the thickness of the top hBN (\SI{59}{nm}) is comparable to the lateral dimension of the split gate (150 nm), the channel region is tuned not only by the channel gate and the back gate, but also by the split gates due to fringe fields. We perform two-terminal transport measurements
at a temperature of \SI{24}{mK}.

To characterize the twist angle between the graphene layers of the sample, we apply a DC current ($I=\SI{10}{nA}$) and measure the resulting voltage drop $V$. We sweep the back gate voltage $V_\textrm{bg}$ while keeping the split gate voltage $V_\textrm{sg}$ and the channel gate voltage $V_\textrm{cg}$ at zero, thus probing the bulk (leads, split-gated region, and channel together). 
The resulting resistance $R=V/I$ is shown in Fig. \ref{Sample}(b), exhibiting several peaks originating from different filling factors $\nu$ of the moir\'e unit cell . 
The peak at around $V_\textrm{bg}=0$ indicates charge neutrality ($\nu=0$). At negative $V_\textrm{bg}$, we observe full filling ($\nu=-1$, where the Fermi level moves into the dispersive band overcoming a small band gap) and half filling ($\nu=-1/2$) peaks for holes. At positive voltages we observe half filling ($\nu=1/2$) for electrons, where the Fermi level resides within the flat bands. Superconductivity appears at filling factors slightly larger than $1/2$. Filling $\nu=1$ for electrons cannot be reached due to the voltage range of the back gate being limited by leakage currents.
We extract the twist angle between the graphene layers of the device from the density of the $\nu=-1$ peak, which leads to \SI{1.1}{\degree} (see Appendix \ref{TwistAngle}).

To demonstrate the area-selective tunability of moir\'e filling factors, Fig.~\ref{Sample}(c) shows $R$ as a function of $V_\textrm{sg}$ and $V_\textrm{bg}$. Filling factors $\nu'=-1, -1/2, 0, 1/2$ and 1 below the split-gates are indicated with black dashed lines.
The inset in Fig.~\ref{Sample}(c) shows a typical current-voltage ($I$--$V$) characteristic in the superconducting regime at $V_\textrm{bg}=\SI{-7.44}{V}$, $V_\textrm{sg}=0$, and $V_\textrm{cg}=0$. Here, a constant contact resistance of \SI{10}{k \ohm}, observed as the offset resistance at zero-bias current in the superconducting regime, is subtracted from the raw data. In the remainder of this manuscript, we consider transport to be superconducting when such an $I$--$V$ characteristic is observed.

A question arising from the map shown in Fig.~\ref{Sample}(c) is the origin of the highly resistive area that spans out in the $\nu'<-1$ region (see the blue region delimited by the dashed pink line). In a standard back-gated MATBG without any top-gates \cite{cao_unconventional_2018,yankowitz_tuning_2019}, only a resistance peak can be observed as a function of $V_\textrm{bg}$ and not a resistive area as a function of $V_\textrm{bg}$ and $V_\textrm{sg}$. In addition, at first glance, one would expect the edge of this high resistive area (dashed pink line) to coincide with the predicted $\nu'=-1$ (labelled dashed black line). The explanation for why this is not the case lies in the next sections. 

\begin{figure*}[ht]
\centering
\includegraphics[scale=1]{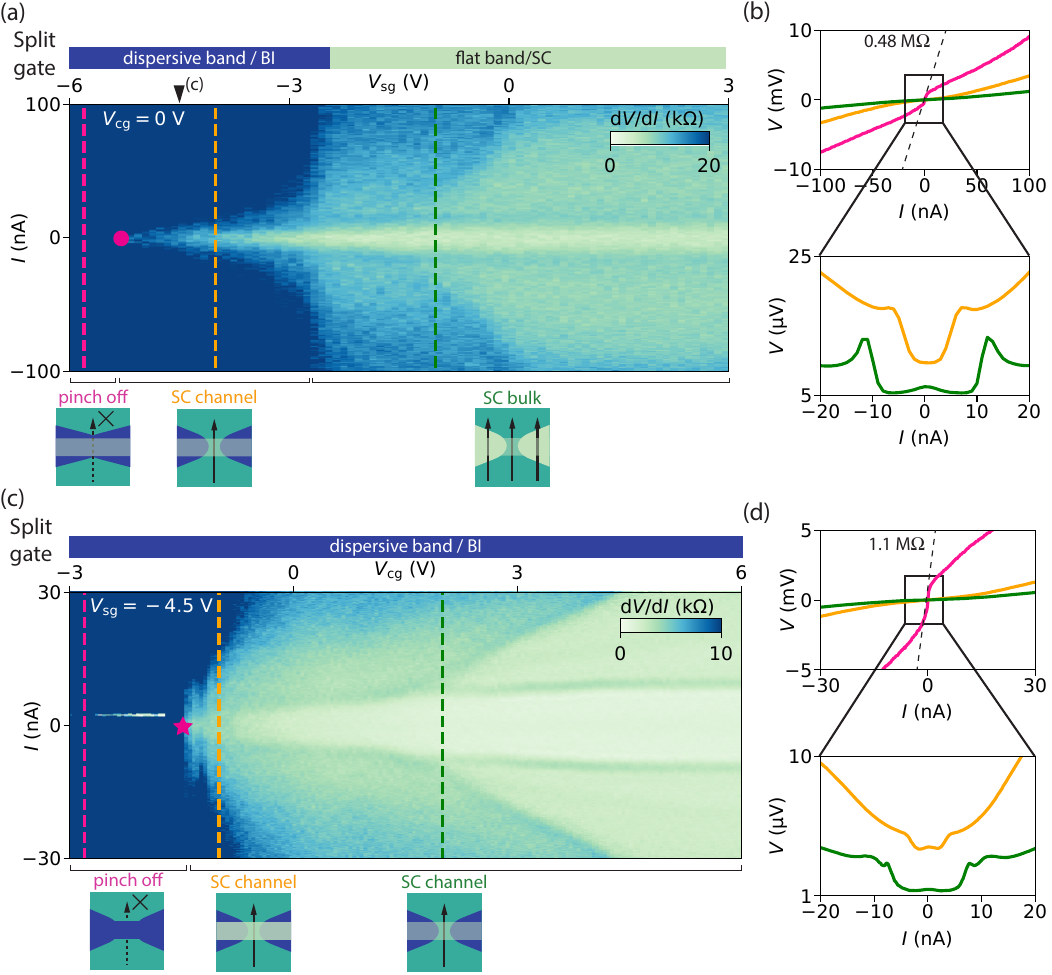}
\caption{Differential resistance (d$V$/d$I$) as a function of current $I$ and gate voltages. (a) Split gate $V_\textrm{sg}$ is swept while keeping $V_\textrm{cg}=\SI{0}{V}$. (b) $IV$ trace and zoom-in picture of d$V$/d$I$ at $V_\textrm{sg}=$-5.8 (black), -4 (yellow), -1 (green) V.(c) Channel gate $V_\textrm{cg}$ is swept while keeping $V_\textrm{sg}=\SI{-4.5}{V}$. (d) $IV$ trace and zoom-in picture of d$V$/d$I$ at $V_\textrm{cg}=$-2.8, -1, and 2 V. The states under the split gate (dispersive band/BI or flat band/SC) are indicated by the bars above the top axes. In all cases, $V_\textrm{bg}$ is kept at zero. Pink, yellow, and green dashed vertical lines show the gate voltage at which the line cuts are taken. Illustrations of the device's state (pinch off, SC channel, or SC bulk) are also shown. Red circle and star symbols show the onset of pinch-off.}
\label{SGCG}
\end{figure*}

\section{Tunable superconducting channel}
\subsection{Control via split gates}
We now study the effect of the split gates on the supercurrent. We tune the leads to the superconducting regime by setting $V_\textrm{bg}=\SI{-7.44}{V}$ [c.f. red arrow in Fig.~\ref{Sample}(c)] and keep $V_\textrm{cg}=0$.
Figure \ref{SGCG}(a) 
shows the d$V$/d$I$ characteristics numerically calculated from the measured dc $V$--$I$ data as a function of the current $I$ and the split-gate voltage $V_\textrm{sg}$.
We assign the state of the carrier gas below the split gates as a function of $V_\textrm{sg}$ according to Fig.~\ref{Sample}(c) and a capacitance model (see Appendix \ref{CapacitanceModel}) as indicated in Fig.~\ref{SGCG}(a) by the colored bar above the color plot.
When $\SI{-6}{V}<V_\textrm{sg}<\SI{-2.8}{V}$, the filling factor under the split gate is smaller than $-1$, and thus the carrier gas in the region below the split gates is either a band insulator (BI) or the Fermi energy is in the dispersive band.
When the voltage is increased to $V_\textrm{sg}>\SI{-2,8}{V}$, the Fermi energy in the region below the split gates is in the flat bands.

In the range $\SI{-6}{V}<V_\textrm{sg}<\SI{-5.5}{V}$ [before the pink dot in Fig. \ref{SGCG}(a)] , the device shows a large zero-bias resistance of $\SI{0.48}{M\Omega}$. This can also be seen in the large slope of the $I$--$V$ characteristic in Fig. \ref{SGCG}(b) (black curve). Such a high resistance despite superconducting leads means, that not only the region below the split gates is resistive, but the resistive region extends into the channel region between them due to fringe fields.

When the split gate voltage is increased to $\SI{-5.5}{V}<V_\textrm{sg}<\SI{-2.8}{V}$ [beyond pink dot in Fig. \ref{SGCG}(a)], the device shows low resistance at low bias and a nonlinear $I$--$V$ characteristic reminiscent of a superconductor as shown by the yellow curve in Fig. \ref{SGCG}(b). 
In this regime, the transport characteristic across the sample remains superconducting even though the area below the split gates is resistive. This indicates that a supercurrent flows through the channel while the split-gated region confines the flow.
When $V_\textrm{sg}$ is increased beyond $\SI{-2.8}{V}$, the nonlinear transport and the critical currents become more prominent [see green curve in Fig.~\ref{SGCG}(b)]. In this regime, the supercurrent extends from the channel into the split-gated regions, eventually rendering the entire device superconducting. 


\subsection{Control via channel gate}
In Figs.~\ref{SGCG}(c) and (d) we show that the conductance of the superconducting channel can also be tuned with the channel gate voltage. It is only the channel that is tuned by $V_\textrm{cg}$ because the metallic split-gate screens the electric field of the channel gate.
We keep the leads superconducting ($V_\textrm{bg}=\SI{-7.44}{V}$) and tune the Fermi energy in the split-gated regions into the dispersive bands ($V_\textrm{sg}=\SI{-4.5}{V}$). This corresponds to the condition in Fig. \ref{SGCG}(a) indicated by the black triangle on the upper axis.


Under these conditions, we see in Fig.~\ref{SGCG}(c) at $V_\textrm{cg}=0$ nonlinear transport indicating superconducting behavior consistent with Fig. ~\ref{SGCG}(a).
By reducing $V_\textrm{cg}$ below $\SI{-1.5}{V}$, thereby tuning the Fermi energy in the channel towards the BI or the dispersive band, the device exhibits a high resistance of $\SI{1.1}{M\Omega}$ and a nonlinear $I$--$V$ characteristic with a large slope around zero bias as shown in Fig. ~\ref{SGCG}(d) (black curve). This means that the superconducting channel is pinched off. Note that the light green line in this area is an artifact.
When $V_\textrm{cg}$ is increased above $\SI{0}{V}$, the superconductivity in the channel is enhanced as illustrated by the green curve in Fig.~\ref{SGCG}(d).
The confinement of the superconducting current to the channel and the local gate-tunability of the channel from the superconducting to the highly resistive (pinch-off) state of the constriction are the central experimental findings of this paper.

\subsection{$V_\textrm{cg}$ vs. $V_\textrm{sg}$ mapping of resistance}
\begin{figure}[h]
\centering
\includegraphics[scale=1]{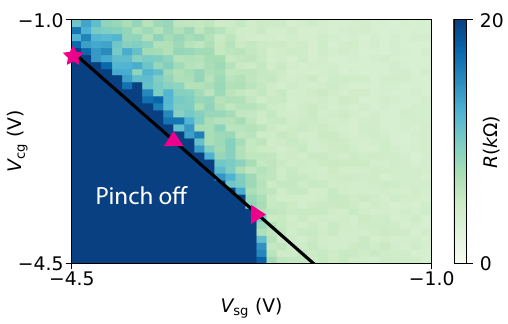}
\caption{Mapping of resistance as a function of $V_\textrm{sg}$ and $V_\textrm{cg}$ at a constant back gate voltage of $V_\textrm{bg}$=-7.44 V. Colored symbols and black line interpolating between them indicate the threshold gate voltages of the pinch off. The gate-voltage regime where high resistances ($R>20 k\Omega$)  are observed is annotated as "Pinch off".}
\label{Pinchoffdiagram}
\end{figure}

To further highlight that the pinch-off in Fig.~\ref{SGCG}(c) originates from the channel, we show the resistance of the device as a function of $V_\textrm{sg}$ and $V_\textrm{cg}$ in Fig.~\ref{Pinchoffdiagram}.
Pink symbols indicate the gate voltages at which the channel is pinched off as observed in the data of Figs.~\ref{SGCG} and \ref{PinchoffExtendedFigure} (see Appendix \ref{VsgVcgMap}). These values lie on a line with a negative slope (see the black line in the Figure).
This means that there exists a regime in which the channel conductance is controlled by both $V_\textrm{sg}$ and $V_\textrm{cg}$.
This is due to the fringe fields of the split-gates, such that the channel conductance is controlled by all three gates. This additionally explains the origin of the discrepancy between the expected $\nu'=1$ (labelled dashed black line) and the high resistance edge (dashed pink line) in Fig.~\ref{Sample}(c). In between the two dashed lines, the resistance is low because even if the split-gated region is resistive, a supercurrent can still flow through the channel. Only when the channel is pinched off due to the effect of the fringing fields we measure a high resistance.

\subsection{Discussion}
In narrow superconducting channels, it has been predicted that the critical current becomes quantized as the channel size is reduced when the BCS coherence length of the superconductor is longer than the channel length \cite{Beenakker1991-dj}. A step-wise change of supercurrent, if not the predicted exact quantization, has been observed in various materials such as InAs \cite{Bauch2005-ii,Abay2013-st, Irie2014, Takanayagi1995}, Ge/Si \cite{Xiang2006-un,Hendrickx2019-kw}, and SrTiO$_{3}$\cite{Mikheev2021-da}.
In our device, we did not observe such quantization of the critical current.
In MATBG, the BCS coherence length is estimated to be from a few tens of nm up to 100 nm \cite{de_vries_gate-defined_2021,Iwakiri2023-gl}. The length of our channel is \SI{400}{nm}, which is much longer than the estimated coherence length. In order to observe the quantization of the critical current, it is necessary to reduce the channel length down to less than \SI{100}{nm}, which is possible by advanced electron beam lithography.

Furthermore, we discuss the conditions at which the pinch-off occurs. In Figs. \ref{SGCG}(a) and (c), the points at which the channel is pinched-off are ($V_\textrm{bg}$, $V_\textrm{sg}$, $V_\textrm{cg}$) = (-7.44, -5.3, 0) and (-7.44, -4.5, -1.5) V, respectively.
Using the capacitance model (see Appendix \ref{CapacitanceModel}), we estimate the corresponding carrier densities $n_\textrm{ch}$ in the channel to be $\SI{-2.23e12}{cm^{-2}}$ and $\SI{-2.56e12}{cm^{-2}}$. For simplicity, we do not take fringe field effects into account in the model. By assuming an isotropic Fermi surface and four-fold degeneracy, the corresponding Fermi wavelengths $\lambda_\textrm{F}=2\sqrt{\pi/|n_\textrm{ch}|}$ are $\SI{23.7}{nm}$ and $\SI{22.1}{nm}$, respectively.
Therefore, one can estimate the minimal width of the channel that supports normal conducting transport to be around this value ($\sim\SI{20}{nm}$), which is significantly smaller than its lithographic value (150 nm). Moreover, in this narrow-channel limit, one can expect conductance quantization in the normal conducting regime. However, we did not observe any signature of normal conductance quantization, presumably due to disorder and the relatively low resistance of the insulating state in MATBG ($\sim\SI{1}{\mega \ohm}$ in this device) compared to gapped Bernal bilayer graphene ($\sim\SI{100}{\mega \ohm}$ \cite{HiskeOverweg}). Nevertheless, the superconducting channel can be formed as long as its resistance, which is ideally zero, is low enough compared to the resistance of the split-gated area.

\section{Conclusion}
We have realized a gate-defined superconducting channel in MATBG by implementing a device with a back gate and two layers of top gates (split gates and channel gate). 
We observe a transition from superconducting to highly resistive (up to $\sim\SI{1}{\mega \ohm}$) transport through the channel by tuning the split-gate and channel gate voltages. This shows that it is possible to define a narrow superconducting channel by tuning the width of the confining resistive areas. The threshold at which the pinch-off occurs depends not only on the channel gate but also on the split gate voltage, indicating the essential role of the fringe field effects.
Our findings serve as a foundation for developing quantum constriction devices like superconducting quantum point contacts, quantum dots and Cooper-pair boxes in MATBG.

\begin{acknowledgments}
We are grateful for fruitful discussions and technical support from Peter Maerki, Thomas Baehler, Rebekka Garreis, Chuyao Tong, Benedikt Kratochwil, Wister Huang, and the ETH FIRST cleanroom facility staff. We acknowledge financial support by the European Graphene Flagship Core3 Project, H2020 European Research Council (ERC) Synergy Grant under Grant Agreement 951541, the European Union’s Horizon 2020 research and innovation program under grant agreement number 862660/QUANTUM E LEAPS, the European Innovation Council under grant agreement number 101046231/FantastiCOF, NCCR QSIT (Swiss National Science Foundation, grant number 51NF40-185902).
K.W. and T.T. acknowledge support from the JSPS KAKENHI (Grant Numbers 21H05233 and 23H02052) and World Premier International Research Center Initiative (WPI), MEXT, Japan.
\end{acknowledgments}

\setcounter{equation}{0}
\setcounter{figure}{0}
\renewcommand{\theequation}{A.\arabic{equation}}
\renewcommand{\thefigure}{A.\arabic{figure}}

\revappendix
\section{Extraction of twist angle} \label{TwistAngle}
We extract the twist angle of the sample using the relation $\theta = 2 \arcsin \left( \frac{a}{2L} \right)$. Here, $a$ is the lattice constant of graphene, and $L$ is the moir\'e periodicity, representing the distance between two AA-stacked regions. $L$ is related to the area $\mathcal{A}$ of the moir\'e unit cell via $L = 2 \sqrt{2 \mathcal{A}/\sqrt{3}}$. 
A moir\'e unit cell can host four electrons due to spin and valley degeneracy. Using this, the full filling resistance peak appears at the density that corresponds to the occupation of 4 electrons per moir\'e unit cell $\mathcal{A} = \frac{4}{n_{\nu=1}}$.

\section{Capacitance model} \label{CapacitanceModel}
To estimate the carrier density in the device, we estimate the capacitance per unit area of the back gate, split gate, and channel gate to be $C_\textrm{bg} = \varepsilon_0 \varepsilon_\textrm{hBN} / d_\textrm{bot}$, $C_\textrm{sg} = \varepsilon_0 \varepsilon_\textrm{hBN} / d_\textrm{top}$, and $C_\textrm{cg} = \varepsilon_0 \varepsilon_\textrm{hBN} \varepsilon_\textrm{AlOx} / (\varepsilon_\textrm{hBN} d_\textrm{top} + \varepsilon_\textrm{AlOx} d_\textrm{AlOx})$. 
Here, $\varepsilon_0=8.854\times10^{-12}$ is the vacuum permittivity, $\varepsilon_\textrm{hBN} = 3.3$ and $\varepsilon_\textrm{AlOx} = 9.5$ are the relative permittivities of the hBN and the aluminium oxide, $d_\textrm{top}$ and $d_\textrm{bot}$ are the thicknesses of the top and bottom hBN and $d_\textrm{AlOx}$ is the thickness of the aluminium oxide layer. 
We then calculate the carrier density of the lead as $n_\textrm{bg} = C_\textrm{bg} V_\textrm{bg} / e$, the region below the split gate as $n_\textrm{sg} = \left(C_\textrm{bg} V_\textrm{bg} + C_\textrm{sg} V_\textrm{sg}\right) / e$, and the region below the channel gate as $n_\textrm{cg} = \left(C_\textrm{bg} V_\textrm{bg} + C_\textrm{cg} V_\textrm{cg}\right) / e$, where $e$ is the elementary charge.

\section{Extended map of $R(V_\textrm{cg}$, $V_\textrm{sg})$} \label{VsgVcgMap}

\begin{figure}[h]
\centering
\includegraphics[scale=0.9]{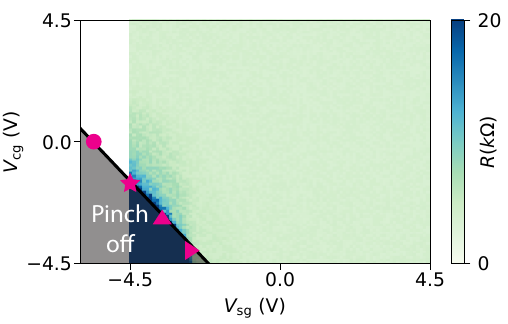}
\caption{Resistance as a function of $V_\textrm{sg}$ and $V_\textrm{cg}$. at a constant back gate voltage of $V_\textrm{bg}$=\SI{-7.44}{V} with an extended plot area.}
\label{PinchoffdiagramFull}
\end{figure}

The same map as Fig.~\ref{Pinchoffdiagram} but with an extended range is shown in Fig.~\ref{PinchoffdiagramFull}. The threshold for pinch-off extracted from the $I$--$V$ characteristic measurements are plotted as symbols. Even though the data cannot be measured due to the leakage of the gate that happened at the last stage of the entire experiment, one can clearly see the linear relation between the threshold value of $V_\textrm{cg}$ and $V_\textrm{sg}$. 

\section{Extended map for $dV/dI(I, V_\textrm{cg}$) and $dV/dI(I, V_\textrm{sg})$} \label{PinchoffExtended}

\begin{figure}[h]
\centering
\includegraphics[scale=0.9]{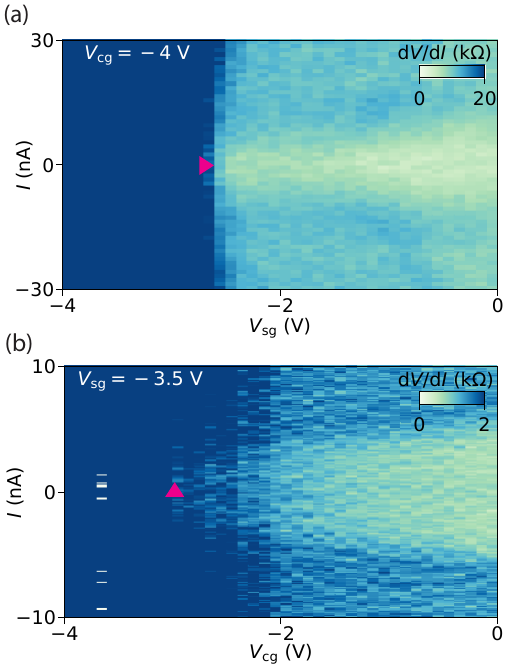}
\caption{d$V$/d$I$ as a function of $V_\textrm{sg}$ with fixed $V_\textrm{bg}=\SI{-7.44}{V}$ and $V_\textrm{cg}=\SI{-4}{V}$. Here, the pinch-off of the channel is observed at $V_\textrm{sg}=\SI{-2.7}{V}$ (see the pink rotated triangular symbol). Figure \ref{PinchoffExtendedFigure}(b) shows d$V$/d$I$ as a function of $V_\textrm{cg}$ with fixed $V_\textrm{bg}=\SI{-7.44}{V}$ and $V_\textrm{sg}=\SI{-3.5}{V}$.}
\label{PinchoffExtendedFigure}
\end{figure}

We measure the behavior of the superconducting channel in the same manner as in Fig.~\ref{SGCG} but in two more configurations as shown in Figure \ref{PinchoffExtendedFigure}.
In Fig. \ref{PinchoffExtendedFigure}(a), $V_\textrm{sg}$ is swept with fixed $V_\textrm{bg}=\SI{-7.44}{V}$ and $V_\textrm{cg}=\SI{-4}{V}$. Here, the pinch-off of the channel is observed at $V_\textrm{sg}=\SI{-2.7}{V}$ (see the pink rotated triangular symbol). 
In Fig. \ref{PinchoffExtendedFigure}(b), $V_\textrm{cg}$ is swept with fixed $V_\textrm{bg}=\SI{-7.44}{V}$ and $V_\textrm{sg}=\SI{-3.5}{V}$. Here, the pinch-off of the channel is observed at $V_\textrm{cg}=\SI{-2.88}{V}$ (see the pink rotated triangular symbol). These points are plotted in Fig. \ref{Pinchoffdiagram} and Fig. \ref{PinchoffdiagramFull}.

\clearpage
\newpage

\bibliography{0_Bibliography}

\end{document}